\documentclass[amssymb,aps,eqsecnum]{revtex4}
\usepackage{nopageno}
\voffset= + 0.96in
\hoffset= + 0.15in
\usepackage{amsmath}
\usepackage{enumerate}
\usepackage{graphicx}
\DeclareGraphicsExtensions{.pdf,.png,.jpg}
\begin{document}
\title{The total energy--momentum tensor for electromagnetic fields in a dielectric}
\author{Michael E. Crenshaw}
\vskip 0.1in
\affiliation{US Army Aviation and Missile Research, Development, and
Engineering Center, Redstone Arsenal, AL 35898, USA}
\centerline{Proc. SPIE, Vol. 10347, Optical Trapping and Optical Micromanipulation XIV; 103473B (2017); doi: 10.1117/12}
\begin{abstract}
There are various formulations of energy--momentum tensors for an
electromagnetic field in a linear dielectric.
The total energy--momentum tensor, comprised of electromagnetic and
material components, must be unique.
We discuss the construction of the total energy--momentum tensor and the
associated conservation laws.
\end{abstract}
\maketitle
\par
\section{Introduction}
\par
Radiation pressure is an observable consequence of optically induced
forces on materials.
On cosmic scales, radiation pressure is responsible for the bending of
the tails of comets as they pass near the sun.
At a much smaller scale, optically induced forces are being investigated
as part of a toolkit for micromanipulation and nanofabrication
technology \cite{BIToolkit}.
A number of practical applications of the mechanical effects of
light--matter interaction are discussed
by Qiu, {\it et al.} \cite{BIQ}.
The promise of the nascent nanophotonic technology for manufacturing
small, low-power, high-sensitivity sensors and other devices has likely
motivated the substantial current interest in optical manipulation
of materials at the nanoscale, see, for example, Ref.~\cite{BIQ} and
the references therein.
While substantial progress toward optical micromanipulation has been
achieved, {\it e.g.} optical tweezers \cite{BIToolkit}, in this report
we limit our consideration to the particular issue of optically induced
forces on a transparent dielectric material.
As a matter of electromagnetic theory, these forces remain indeterminate
and controversial.
Due to the potential applications in nanotechnology, the century-old
debate regarding these forces, and the associated momentums, has ramped
up considerably in the physics community.
\vskip 0.125in
\par
The energy--momentum tensor is the centerpiece of conservation laws for
the unimpeded, inviscid, incompressible flow of non-interacting
particles in the continuum limit in an otherwise empty volume.
The foundations of the energy--momentum tensor and the associated tensor
conservation theory come to electrodynamics from classical continuum
dynamics by applying the divergence theorem to a Taylor series expansion
of a property density field of a continuous flow in an otherwise empty
volume.
The dust tensor is a particularly simple example of an energy--momentum
tensor that deals with particles of matter in the continuum limit in
terms of the mass density $\rho_m$, energy density $\rho_m c^2$, and
momentum density $\rho_m {\bf v}$.
Newtonian fluids can behave very much like dust with the same
energy--momentum tensor.
The energy and momentum conservation properties of light propagating in
the vacuum were long-ago cast in the energy--momentum tensor formalism
in terms of the electromagnetic energy density and electromagnetic
momentum density.
However, extrapolating the tensor theory of energy--momentum
conservation for propagation of light in the vacuum to propagation of
light in a simple linear dielectric medium has proven to be problematic
and controversial.
A dielectric medium is not "otherwise empty" and it is typically assumed
that optically induced forces accelerate and decelerate nanoscopic
material constituents of the dielectric.
The corresponding material energy--momentum tensor is added to the
electromagnetic energy--momentum tensor to form the total
energy--momentum tensor, thereby ensuring that the total energy and the
total momentum of the thermodynamically closed system remain constant
in time.
\par
\section{The Total Energy--Momentum tensor}
\par
The total energy--momentum tensor for the flow of light in a linear 
medium is the sum of the electromagnetic energy--momentum tensor
and a material energy--momentum tensor.
The typical development \cite{BIPfei,BIRL,BIObuk}, reviewed by
Pfeifer, Nieminen, Heckenberg, and Rubinsztein-Dunlop \cite{BIPfei},
shows
\begin{equation}
T^{\alpha\beta}_{\rm EM,Abr}= \left [
\begin{matrix}
\frac{1}{2}\left ( {\bf D}\cdot{\bf E}+{\bf H}\cdot{\bf B} \right )
& {\bf E}\times{\bf H}
\cr
{\bf E}\times{\bf H}
& -{\bf E}\wedge{\bf D} -{\bf H}\wedge{\bf B}
+ \frac{1}{2}
\left ({\bf E}\cdot{\bf D}+{\bf H}\cdot{\bf B}\right ){\bf I}
\cr
\end{matrix}
\right ] ,
\label{EQf2.01}
\end{equation}
in Heaviside--Lorentz units and
\begin{equation}
T^{\alpha\beta}_{\rm mat,Abr}= \left [
\begin{matrix}
\rho_m c^2
&\rho_mc{\bf v}
\cr
\rho_mc{\bf v}
&\rho_m{\bf v}\wedge {\bf v}
\cr
\end{matrix}
\right ] \, .
\label{EQf2.02}
\end{equation}
Here, $T^{\alpha\beta}_{\rm EM,Abr}$ is the well-known Abraham
electromagnetic energy--momentum tensor,
$T^{\alpha\beta}_{\rm mat,Abr}$ is the well-known dust tensor,
and $\rho_m$ is the mass density of the dust.
The total energy--momentum tensor, $T^{\alpha\beta}_{\rm total}=
T^{\alpha\beta}_{\rm EM,Abr} +T^{\alpha\beta}_{\rm mat,Abr}$ is
\begin{equation}
T^{\alpha\beta}_{\rm total}= \left [
\begin{matrix}
\frac{1}{2}\left ( {\bf D}\cdot{\bf E}+{\bf H}\cdot{\bf B} \right )
+\rho_mc^2
&{\bf E}\times{\bf H} +\rho_mc {\bf v}
\cr
\frac{1}{c}\left ( {\bf E}\times{\bf H}\right ) +\rho_mc {\bf v}
& -{\bf E}\wedge{\bf D} -{\bf H}\wedge{\bf B}
+ \frac{1}{2}
\left ({\bf E}\cdot{\bf D}+{\bf H}\cdot{\bf B}\right ){\bf I}
+\rho_m{\bf v}\wedge {\bf v}
\cr
\end{matrix}
\right ] \, .
\label{EQf2.03}
\end{equation}
\vskip 0.125in
\par
According to the scientific literature \cite{BIPfei,BIObuk}, the
total energy--momentum tensor can also be constructed from the
sum of the Minkowski electromagnetic energy--momentum tensor
\begin{equation}
T^{\alpha\beta}_{\rm EM,Mink}= \left [
\begin{matrix}
\frac{1}{2}\left ( {\bf D}\cdot{\bf E}+{\bf H}\cdot{\bf B} \right )
&{\bf E}\times{\bf H}
\cr
{\bf D}\times{\bf B}
& -{\bf E}\wedge{\bf D} -{\bf H}\wedge{\bf B}
+ \frac{1}{2}
\left ({\bf E}\cdot{\bf D}+{\bf H}\cdot{\bf B}\right ){\bf I}
\cr
\end{matrix}
\right ] 
\label{EQf2.04}
\end{equation}
and an appropriate material tensor $T^{\alpha\beta}_{\rm mat,Mink}$.
However, there is no physical model for this material tensor
as there is for $T^{\alpha\beta}_{\rm mat,Abr}$.
Instead, the material tensor \cite{BIPfei,BIObuk}
\begin{equation}
T^{\alpha\beta}_{\rm mat,Mink}= \left [
\begin{matrix}
\rho_m c^2
&\rho_mc{\bf v}
\cr
\rho_mc{\bf v}
+{\bf E}\times{\bf H}
- {\bf D}\times{\bf B}
&\rho_m{\bf v}\wedge {\bf v}
\cr
\end{matrix}
\right ]
\label{EQf2.05}
\end{equation}
that accompanies the Minkowski electromagnetic energy--momentum tensor,
Eq.~(\ref{EQf2.04}), is obtained phenomenologically by starting with the
total energy--momentum tensor, Eq.~(\ref{EQf2.03}), then subtracting the
Minkowski electromagnetic energy--momentum tensor, Eq.~(\ref{EQf2.04}).
\vskip 0.125in
\par
The total energy and total linear momentum are constrained by the
conservation law \cite{BIGold}
\begin{equation}
\frac{\partial T_{\rm total}^{\alpha\beta}}{\partial x^{\beta}}=0 \, ,
\label{EQf2.06}
\end{equation}
where $x^{\beta}\in \{ x^0=ct,x^1=x,x^2=y,x^3=z \}$.
Substituting Eq.~(\ref{EQf2.03}) into Eq.~(\ref{EQf2.06}), one obtains
\begin{equation}
\frac{1}{c}\frac{\partial}{\partial t}\left ( \rho_e+\rho_m c^2\right )
+\nabla\cdot \left ( {\bf E}\times{\bf H}+\rho_m c {\bf v}\right )=0
\label{EQf2.07}
\end{equation}
for the $\alpha=0$ element.
Here, $\rho_e$ denotes the electromagnetic energy density
\begin{equation}
\rho_e=
\frac{1}{2}\left ( {\bf E}\cdot{\bf D}+{\bf H}\cdot{\bf B} \right ) \, .
\label{EQf2.08}
\end{equation}
An additional constraint on conservation is that the total energy and
the total momentum must remain constant as a wave packet transits
from free-space into the material.
For the total energy and total linear momentum, this constraint is
\begin{equation}
P^{\alpha}(t)=\int_{\sigma} T^{\alpha 0}_{\rm total} dv=P^{\alpha}(t_0)
\label{EQf2.09}
\end{equation}
where the volume of integration has been extended to
all-space, $\sigma$.
For the conservation of total angular momentum, the constraint
\begin{equation}
{T}^{\alpha \beta}_{\rm total} = {T}^{\beta \alpha}_{\rm total} 
\label{EQf2.10}
\end{equation}
is often employed although the requirement that the total angular
momentum is constant in time does not necessarily require the total
energy--momentum tensor to be diagonally symmetric \cite{BIObuk}.
Some authors take this caveat to mean that that the specific relationship
between ${T}^{\alpha \beta}_{\rm total}$ and its transpose is immaterial
because it is what it needs to be to insure that total angular momentum
is conserved in a thermodynamically closed system.
There is, however, a relationship, although is is more complicated than
Eq.~(\ref{EQf2.10}).
In order to avoid unnecessary complications, we posit the case of ordinary
unstructured fields in the common plane-wave limit incident on the
dielectric medium from the vacuum of free space.
Then, then the total energy--momentum tensor of the transmitted field can be
taken as diagonally symmetric if the total energy--momentum tensor of the
incident field, from vacuum, is diagonally symmetric.
\vskip 0.125in
\par
Adopting the constraint on the total linear momentum,
${\bf P}_{\rm total}=( P_{\rm total}^1,P_{\rm total}^2,P_{\rm total}^3)$,
Pfeifer, Nieminen, Heckenberg, and Rubinsztein-Dunlop \cite{BIPfei}
derive 
\begin{equation}
\rho_m c{\bf v}=(n-1){\bf E}\times{\bf H} 
\label{EQf2.11}
\end{equation}
by subtracting the electromagnetic momentum from the total momentum,
where the latter is determined from Eq.~(\ref{EQf2.09}) taking into
account the well-known change in amplitude and spatial width of 
electromagnetic fields in a dielectric.
Taking the mass density of the material $\rho_m$ to be constant for
a quasimonochromatic/monochromatic field in the plane wave limit and
substituting Eq.~(\ref{EQf2.11}) into Eq.~(\ref{EQf2.07}), we find that
\begin{equation}
\frac{1}{c}\frac{\partial\rho_e}{\partial t}
+\nabla\cdot
(n{\bf E}\times{\bf H}) =0 \, .
\label{EQf2.12}
\end{equation}
This result is manifestly false for a quasimonochromatic field because
the two non-zero terms depend on different powers of the independent
parameter $n$.
To see this, one can perform the calculus operations on fields written
in terms of a constant field amplitude and a carrier wave,
$e^{-(i\omega t \pm n(\omega/c)z)}$, as one does when deriving the
Fresnel relations, or one can simply note that Eq.~(\ref{EQf2.12}) is
incommensurate with the Poynting theorem (Eq.~(\ref{EQf3.02}), below).
Consequently, we are forced to argue that $\rho_m=(n-1)\rho_e$
oscillates at optical frequencies such that Eq.~(\ref{EQf2.07}) is
commensurate with the Poynting theorem, but at the expense of violating
the conservation law, Eq.~(\ref{EQf2.06}), which is violated because the
timelike coordinate, $ct$, becomes index-dependent, $ct/n$, if
one follows this line of reasoning.
\par
\section{Maxwellian continuum electrodynamics}
\par
Having determined that the standard model of the total energy--momentum
tensor produces a total energy conservation law that is false, we would
like to circumscribe the problem.
We start with the derivation of the electromagnetic energy--momentum
tensor.
The familiar Maxwell--Minkowski equations can be written
as \cite{BIJack}
\begin{subequations}
\begin{equation}
\nabla\times{\bf H}-\frac{1}{c}\frac{\partial{\bf D}}{\partial t}=0
\label{EQf3.01a}
\end{equation}
\begin{equation}
\nabla\cdot{\bf B}=0
\label{EQf3.01b}
\end{equation}
\begin{equation}
\nabla\times{\bf E}+\frac{1}{c}\frac{\partial{\bf B}}{\partial t}=0
\label{EQf3.01c}
\end{equation}
\begin{equation}
\nabla\cdot{\bf D}=0 \, ,
\label{EQf3.01d}
\end{equation}
\label{EQf3.01}
\end{subequations}
where the macroscopic fields, ${\bf E}$, ${\bf D}$, ${\bf B}$,
and ${\bf H}$, are functions of position, ${\bf r}$, and time, $t$.
As is commonly done, we limit consideration to simple linear dielectric
media in which the center frequency of the exciting
quasimonochromatic/monochromatic field is away from material resonances.
In this regime, absorption and frequency dispersion can be treated
as negligible in the lowest order of approximation.
Then, the refractive index, $n({\bf r})$, depends on the center
frequency of the exciting quasimonochromatic/monochromatic field but is
otherwise a real, time-independent, single-valued function of position
In this regime, we can freely use the familiar constitutive relations,
${\bf D}=n^2{\bf E}$ and ${\bf B}={\bf H}$,
for a simple linear dielectric.
Poynting's theorem
\begin{equation}
\frac{1}{c}\frac{\partial \rho_e }{\partial t}+
\nabla\cdot\left ( {\bf E}\times{\bf H}\right ) =0
\label{EQf3.02}
\end{equation}
is derived by subtracting the scalar product of Eq.~(\ref{EQf3.01a})
with ${\bf E}$ from the scalar product of Eq.~(\ref{EQf3.01c}) with
${\bf H}$ and applying common vector identities.
Likewise,
\begin{equation}
\frac{1}{c}\frac{\partial}{\partial t}
\left ( {\bf D}\times{\bf B}\right )_i
+ \sum_j \frac{\partial}{\partial x_i}{W}^{\rm Mink}_{ij} =
{\bf f}^{\rm Mink}_i
\label{EQf3.03}
\end{equation}
is rigorously derived from the Maxwell--Minkowski field equations,
Eqs.~(\ref{EQf3.01}), by adding the cross-product of
Eq.~(\ref{EQf3.01a}) with ${\bf B}$ to the cross-product of ${\bf D}$
with Eq.~(\ref{EQf3.01c}) and simplifying the result using
Eqs.~(\ref{EQf3.01b}) and (\ref{EQf3.01d}) \cite{BIJack}.
Here,
\begin{equation}
{\bf f}^{\rm Mink}=
({\bf E}\cdot\nabla(n^2)){\bf E}=
{\bf E}\times({\bf E}\times\nabla(n^2))+{\bf E}^2\nabla(n^2)
\label{EQf3.04}
\end{equation}
is the Minkowski force density, which {\it does not}
reduce to $-({1}/{2}){\bf E}^2\nabla(n^2)$ \cite{BIMilBoy}.
In fact, the Minkowski force is zero in the plane-wave limit.
The elements of ${W}^{\rm Mink}$ are the elements of the matrix
\begin{equation}
{W}^{\rm Mink}_{ij}=-D_i E_j -H_i B_j 
+\frac{1}{2}\left ( {\bf D}\cdot {\bf E}+
{\bf H}\cdot {\bf B}\right ) \delta_{ij}. 
\label{EQf3.05}
\end{equation}
Next, we write the scalar equation, Eq.~(\ref{EQf3.02}),
and the three scalar components of the vector equation,
Eq.~(\ref{EQf3.03}), row-wise, as a differential equation
\begin{equation}
\frac{T_{\rm EM,Mink}^{\alpha\beta}}{\partial x^{\beta}}
={\bf f}_{\alpha}^{\rm Mink}
\label{EQf3.06}
\end{equation}
where
${f}_{\alpha}^{\rm Mink}=(0,{\bf f}^{\rm Mink})$ is an element of the
four-force density and
\begin{equation}
T^{\alpha\beta}_{\rm EM,Mink}= \left [
\begin{matrix}
\frac{1}{2}\left ( {\bf E}\cdot{\bf D}+{\bf H}\cdot{\bf B} \right )
&\frac{1}{c}\left ( {\bf E}\times{\bf H}\right )
\cr
\frac{1}{c}\left ( {\bf D}\times{\bf B}\right )
& -{\bf E}\wedge{\bf D} -{\bf H}\wedge{\bf B}
+ \frac{1}{2}
\left ({\bf E}\cdot{\bf D}+{\bf H}\cdot{\bf B}\right ){\bf I}
\cr
\end{matrix}
\right ] ,
\label{EQf3.07}
\end{equation}
is, by definition, a four-by-four square matrix that is historically
known as the Minkowski energy--momentum tensor.
\vskip 0.125in
\par
We can also write the matrix differential equation, Eq.~(\ref{EQf3.06}),
in terms of the Abraham electromagnetic momentum density.
Subtracting the Abraham force density
\begin{equation}
{\bf f}_{\rm Abr}=
\frac{\partial}{\partial t} \frac{(n^2-1){\bf E}\times{\bf H}}{c} 
\label{EQf3.08}
\end{equation}
from both sides of Eq.~(\ref{EQf3.03}), we obtain
\begin{equation}
\frac{1}{c}\frac{\partial}{\partial t}
\left ( {\bf E}\times{\bf H}\right )_i
+ \sum_j\frac{\partial}{\partial x_i} {W}^{\rm Mink}_{ij}
= {\bf f}^{\rm Mink}_i
-\left (\frac{\partial}{\partial t}\frac{(n^2-1){\bf E}
\times{\bf H}}{c}\right )_i \, .
\label{EQf3.09}
\end{equation}
Combining Eq.~(\ref{EQf3.02}) with Eq.~(\ref{EQf3.09}), row-wise,
as before, we obtain a differential equation
\begin{equation}
\partial_{\beta}T_{\rm EM,Abr}^{\alpha\beta}={\bf f}^{\rm Mink}_{\alpha}
- \left (\frac{(n^2-1){\bf E}\times{\bf H}}{c}\right ) _{\alpha} \, .
\label{EQf3.10}
\end{equation}
for each $\alpha$, where $\partial_{\beta}$ is an element of the
four-divergence operator
\begin{equation}
\partial_{\beta}=
\left ( \frac{1}{c}\frac{\partial}{\partial t},
\partial_x,\partial_y,\partial_z \right ) \, .
\label{EQf3.11}
\end{equation}
Many authors claim that the Abraham force density, Eq.~(\ref{EQf3.08}),
that comprises the last term of Eq.~(\ref{EQf3.08}) can be neglected
because the time average of the fluctuating fields is essentially zero.
This is a consequence of analyzing the Abraham force density,
Eq.~(\ref{EQf3.08}), separate from the dynamical equation,
Eq.~(\ref{EQf3.09}). 
If the Abraham force can be neglected, then
\begin{equation}
\sum_j \frac{\partial}{\partial x_j}{W}^{\rm Mink}_{ij} =
 {\bf  f}_{i}^{\rm Mink}
\label{EQf3.12}
\end{equation}
because the first and last terms of Eq.~(\ref{EQf3.09}) differ by
a non-negligible multiplicative constant, $n^2-1$.
Consequently, neglect of the Abraham force results in a manifestly false
static equation for propagating electromagnetic fields.
\par
\section{Lagrangian Equations of Motion}
\par
It is often claimed that Eqs.~(\ref{EQf3.02}) and (\ref{EQf3.03}) are
the electromagnetic energy and momentum conservation laws.
However, the energy and momentum conservation laws originate in
classical continuum dynamics, principally fluid dynamics, and not in
continuum electrodynamics.
Equations~(\ref{EQf3.02}) and (\ref{EQf3.03}) are theorems of the
macroscopic Maxwell--Minkowski equations, Eqs.~(\ref{EQf3.02}), and have
the outward appearance of the conservation law, Eq.~(\ref{EQf2.06}).
However, we have seen that it is impossible to construct a total
energy--momentum tensor from the Maxwell--Minkowski equations that
satisfies the spacetime conservation laws.
Although it is common to introduce an adjustable relationship,
{\it e.g.,} Eq~(\ref{EQf2.11}), between the electromagnetic and material
components of energy and momentum in order to satisfy one of the
conservation laws this action causes the violation of another
conservation law that was previously satisfied.
Specifically, it is not possible to satisfy both Eq.~(\ref{EQf2.06}) and
(\ref{EQf2.07}) simultaneously within the formalism of Maxwellian
continuum electrodynamics.
\vskip 0.125in
\par
The spacetime conservation law, Eq.~(\ref{EQf2.06}), reflects the
conservation of a scalar property in the continuum limit of an unimpeded
inviscid, incompressible flow of non-interacting particles in terms of
the equality of the net rate of flux out of an otherwise empty volume
and the time rate of change of the property density field.
This description perfectly fits the propagation of light in the vacuum.
However, several of the conditions are violated if the light propagates
through a simple linear dielectric medium.
First, the light slows down as it enters the medium indicating some
sort of impediment to free flow.
Second, the volume contains a linear medium and is therefore not 
"otherwise empty".
Third, the volume occupied by the field inside the dielectric is smaller
than the volume occupied by the same field as it is incident from the
vacuum due to the reduced speed of light in a dielectric, violating the
condition of incompressible flow.
\vskip 0.125in
\par
At this point, we turn to Lagrangian field dynamics to derive new field
equations for macroscopic fields in a simple linear dielectric.
The classical Lagrangian is \cite{BIGold,BIJack}
\begin{equation}
L=\frac{1}{2} \int_{\sigma} (T-V) dv \, ,
\label{EQf4.01}
\end{equation}
where $T$ is the kinetic energy density, $V$ is the potential energy
density, and integration is performed over all-space $\sigma$.
We write the classical Lagrangian for macroscopic fields in a linear
dielectric as
\begin{equation}
L=\frac{1}{2} \int_{\sigma} \left (
\left (
\frac{n}{c}\frac{\partial{\bf A}}{\partial t}
\right )^2
- \left ({\nabla\times{\bf A}} \right )^2
\right ) dv \, .
\label{EQf4.02}
\end{equation}
The Lagrangian density,
\begin{equation}
{\cal L}=\frac{1}{2}\left (
\left (
\frac{n}{c}\frac{\partial{\bf A}}{\partial t}
\right )^2
- \left ({\nabla\times{\bf A}} \right )^2
\right ) \, ,
\label{EQf4.03}
\end{equation}
is the integrand of the Lagrangian, Eq.~(\ref{EQf4.02}).
\vskip 0.125in
\par
We consider an arbitrarily large region of space to be filled with an
isotropic homogeneous transparent linear dielectric medium that
is characterized by a linear refractive index $n$.
We limit our attention to simple linear media and we write a new
time-like variable
\begin{equation}
\bar x^0=\frac{ct}{n} \, .
\label{EQf4.04}
\end{equation}
We take the re-parameterized Lagrangian density
\begin{equation}
{\cal L}=\frac{1}{2} \left (
\left (
\frac{\partial{\bf A}}{\partial \bar x^0}
\right )^2
- \left (\nabla\times{\bf A} \right )^2
\right )
\label{EQf4.05}
\end{equation}
as our starting point and apply Lagrangian field theory to
systematically derive equations of motion for the macroscopic fields in
an arbitrarily large isotropic homogenous block of simple linear 
dielectric material.
\vskip 0.125in
\par
The Lagrange equation for electromagnetic fields in the vacuum is
\begin{equation}
\frac{d}{d t}\frac{\partial{\cal L}}
{\partial (\partial A_j /\partial t)}
+\sum_i\frac{\partial}{\partial x_i}
\frac{\partial{\cal L}}{\partial (\partial A_j /\partial x_i)}
=\frac{\partial {\cal L}}{\partial A_j} \, ,
\label{EQf4.06}
\end{equation}
although Eq.~(\ref{EQf4.06}) is also used for electromagnetic fields
in linear media \cite{BICT}.
In terms of the re-parameterized temporal coordinate,
Eq.~(\ref{EQf4.04}), the preceding equation becomes
\begin{equation}
\frac{d}{d \bar x^0}\frac{\partial{\cal L}}
{\partial (\partial A_j /\partial \bar x^0)}
+\sum_i\frac{\partial}{\partial x_i}
\frac{\partial{\cal L}}{\partial (\partial A_j /\partial x_i)}
=\frac{\partial {\cal L}}{\partial A_j}
\label{EQf4.07}
\end{equation}
for simple linear dielectric materials.
Substituting the Lagrangian density, Eq.~(\ref{EQf4.05}), into
Eq.~(\ref{EQf4.07}), we obtain the components
\begin{subequations}
\begin{equation}
\frac{\partial{\cal L}}
{\partial (\partial A_{j}/\partial \bar x^0)}
=\frac{\partial A_j}{\partial \bar x^0}
\label{EQf4.08a}
\end{equation}
\begin{equation}
\frac{\partial \cal L}{\partial A_j}=0
% \frac{nJ_j}{c}
\label{EQf4.08b}
\end{equation}
\begin{equation}
\sum_i\frac{\partial}{\partial  x_i}
\frac{\partial{\cal L}}{(\partial_{i} A_{j}/\partial x_i)}
=[\nabla\times(\nabla\times {\bf A})]_j \, .
\label{EQf4.08c}
\end{equation}
\label{EQf4.08}
\end{subequations}
Substituting components, Eqs.~(\ref{EQf4.08}), into Eq.~(\ref{EQf2.02}),
the Lagrange equations of motion for the electromagnetic field in a
dielectric are the three orthogonal components of the vector wave
equation
\begin{equation}
\nabla\times(\nabla\times {\bf A})
+ \frac{\partial^2{\bf A}}{\partial (\bar x^0)^2} =0 \, .
\label{EQf4.09}
\end{equation}
The second-order equation, Eq.~(\ref{EQf4.09}), can be written as a
set of first-order differential equations and we introduce macroscopic
field variables
\begin{equation}
{\bf \Pi}=
\frac{\partial{\bf A}}{\partial \bar x^0}
\label{EQf4.10}
\end{equation}
\begin{equation}
{\bf B}= \nabla\times{\bf A}
\label{EQf4.11}.
\end{equation}
The macroscopic field variable ${\bf \Pi}$, Eq.~(\ref{EQf4.10}), is
selected by the canonical momentum field density whose
components were defined in Eq.~(\ref{EQf4.08a}).
Note that both of the terms in the Lagrangian density,
${\cal L}={\bf \Pi}^2-{\bf B}^2$, and the Hamiltonian density,
${\cal H}={\bf \Pi}^2+{\bf B}^2$, are quadratic.
\vskip 0.125in
\par
We substitute the definition of the canonical momentum field
${\bf \Pi}$, Eq.~(\ref{EQf4.10}), and the definition of the magnetic
field ${\bf B}$, Eq.~(\ref{EQf4.11}), into the wave equation,
Eq.~(\ref{EQf4.09}), to obtain
\begin{equation}
\nabla\times{\bf B}
+\frac{\partial {\bf \Pi}}{\partial \bar x^0}
=0 \, ,
\label{EQf4.12}
\end{equation}
which is similar to the Maxwell--Amp\`ere law.
Taking the divergence of Eq.~(\ref{EQf4.11}), we obtain
\begin{equation}
\nabla\cdot{\bf B}= 0 \, .
\label{EQf4.13}
\end{equation}
Applying the operator Eq.~(\ref{EQf4.10}) produces
a version of Faraday's Law,
\begin{equation}
\nabla\times{\bf \Pi}
-\frac{\partial{\bf B}}{\partial \bar x^0}
= \frac{\nabla n}{n}\times{\bf \Pi} \, .
\label{EQf4.14}
\end{equation}
Finally,
\begin{equation}
\nabla\cdot{\bf \Pi}=
-\frac{\nabla n}{n}\cdot {\bf \Pi} 
\label{EQf4.15}
\end{equation}
is a modified version of Gauss's law that is obtained by integrating the
divergence of Eq.~(\ref{EQf4.12}) with respect to the temporal
coordinate.
This completes the set of first-order equations of motion for the
macroscopic fields, Eqs.~(\ref{EQf4.12})--(\ref{EQf4.15}).
\vskip 0.125in
\par
The unusual appearance of Eqs.~(\ref{EQf4.12})--(\ref{EQf4.15}) may
cause some readers to question their validity or range of applicability.
Various reviewers of Ref.~\cite{BICrenshaw} have said that the superficially
equivalent equations derived as identities of the
Maxwell--Minkowski equations under a re-parameterization by the
direct substitution of ${\bf \Pi}= - n{\bf E}$ and $\bar x^0=ct/n$
into Eqs.~(\ref{EQf3.01}) are
wrong because:
1) they violate Einstein's special theory of relativity,
2) they violate spacetime conservation laws,
3) they violate conditions of holonomy,
4) they are valid only in some pathological limit, such as a spatially
infinite medium,
5) they were derived using (unspecified) implicit or hidden axioms,
6) there are (unspecified) manifest errors in the derivation.
Other reviewers contend that Eqs.~(\ref{EQf4.12})--(\ref{EQf4.15})
cannot possibly be false because they are identities of the
Maxwell--Minkowski equations, Eqs.~(\ref{EQf3.01}) under a
re-parameterization by the direct substitution of
${\bf \Pi}= - n{\bf E}$ and $\bar x^0=ct/n$.
However, both sets of careful readers are only partially correct.
\vskip 0.125in
\par
Let us briefly examine these misconceptions:
1). Einstein derived special relativity using coordinate transformations
between inertial reference frame in the vacuum of free space.
If a light pulse is emitted from the origin at a time $t=0$ into the
vacuum then spherical wavefronts are defined by
\begin{equation}
x^2+y^2+z^2-(x^0)^2=0
\label{EQf4.16}
\end{equation}
in a flat four-dimensional Minkowski spacetime $S_v(x^0=ct,x,y,z)$.
Von-Laue used the velocity sum rule to derive a theory of special
relativity for a dielectric and Eqs.~(\ref{EQf4.12})--(\ref{EQf4.15})
are proven false by Einstein--von Laue dielectric special relativity.
This should be ``case-closed'' except, in the case of a pulse emitted
into a simple isotropic homogeneous dielectric linear medium, spherical
wavefronts are defined by
\begin{equation}
x^2+y^2+z^2-(\bar x^0)^2=0
\label{EQf4.17}
\end{equation}
in a flat four-dimensional non-Minkowski material spacetime
$S_{d}(\bar x^0,x,y,z)$.
Consequently, a different version of continuum electrodynamics
that is defined on a corresponding material spacetime is associated
with each isotropic homogeneous linear medium.
This result is correlated with Rosen's special relativities in which a
different index-dependent version of special relativity is associated
with each isotropic homogeneous linear dielectric \cite{BIRosen}.
Although Rosen's theory of dielectric special relativities is 
phenomenological in nature, it was derived rigorously in
Ref.~\cite{BICrenshaw} using coordinate transformations between inertial
frames of reference in an arbitrarily large isotropic homogeneous
simple linear dielectric medium in which the speed of light is $c/n$.
Because each value of the refractive index is identified with a
different material spacetime, the rigorous theory is limited
to arbitrarily large isotropic homogeneous simple dielectric
linear media, although piece-wise homogeneous materials can also be
treated using the appropriate boundary conditions.
\vskip 0.125in
\par
2). The spacetime conservation laws were derived for an unimpeded
inviscid, incompressible flow of non-interacting particles in the 
continuum limit through an otherwise empty volume.
The resulting spacetime conservation law, Eq.~(\ref{EQf2.06}),
is clearly violated by Eqs.~(\ref{EQf4.12})--(\ref{EQf4.15}).
However, the premises of the conservation law, Eq.~(\ref{EQf2.06}),
are violated for light propagating into a simple linear dielectric,
as described in the introduction to Sec. IV.
Combining Eqs.~(\ref{EQf4.12})--(\ref{EQf4.15}) as in Sec. 3, we obtain 
the energy conservation law,
\begin{equation}
\frac{\partial \rho_e }{\partial \bar x^0}+
\nabla\cdot\left ( {\bf B}\times{\bf \Pi}\right ) =0 
\label{EQf4.18}
\end{equation}
and the momentum conservation law,
\begin{equation}
\frac{\partial}{\partial x^0}
\left ( {\bf  B}\times{\bf \Pi}\right )_i
+ \sum_j \frac{\partial}{\partial x_j}{W}^{\rm new}_{ij} =
{\bf f}^{\rm new}_i \, ,
\label{EQf4.19}
\end{equation}
where
\begin{equation}
{W}^{\rm new}_{ij}=-{\Pi}_i {\Pi}_j -B_i B_j 
+\frac{1}{2}\left ( {\bf \Pi}\cdot {\bf \Pi}+
{\bf B}\cdot {\bf B}\right ) \delta_{ij} 
\label{EQf4.20}
\end{equation}
and 
\begin{equation}
{\bf f}^{\rm new}_i=
\left (
0,{\bf \Pi}\times\left (\frac{\nabla n}{n}\times{\bf \Pi} \right )
\right ) \, .
\label{EQf4.21}
\end{equation}
Combining Eq.~(\ref{EQf4.18}) with Eq.~(\ref{EQf4.19}), row-wise, as
before, these two conservation laws can be written as a single matrix
differential equation 
\begin{equation}
\bar\partial_{\beta}T_{\rm total}^{\alpha\beta}
={\bf f}^{\rm new}_{\alpha}
\label{EQf4.22}
\end{equation}
for each $\alpha$, where $\bar\partial_{\beta}$ is an element of the
material four-divergence operator
\begin{equation}
\partial_{\beta}= \left ( \frac{\partial}{\partial \bar x^0},
\partial_x,\partial_y,\partial_z \right ) \, .
\label{EQf4.23}
\end{equation}
and
\begin{equation}
T^{\alpha\beta}_{\rm total}= \left [
\begin{matrix}
\frac{1}{2}\left ( {\bf \Pi}\cdot{\bf \Pi}+{\bf B}\cdot{\bf B} \right )
&{\bf B}\times{\bf \Pi}
\cr
{\bf B}\times{\bf \Pi}
& -{\bf \Pi}\wedge{\bf \Pi} -{\bf H}\wedge{\bf B}
+ \frac{1}{2}
\left ({\bf \Pi}\cdot{\bf \Pi}+{\bf B}\cdot{\bf B}\right ){\bf I}
\cr
\end{matrix}
\right ] ,
\label{EQf4.24}
\end{equation}
is the diagonally symmetric total energy--momentum tensor.
Then the conservation law that was derived for a continuous
flow in the vacuum, Eq.~(\ref{EQf2.06}), becomes
\begin{equation}
\frac{\partial T_{total}^{\alpha\beta}}{\partial \bar x^{\beta}}=0 \, ,
\label{EQf4.25}
\end{equation}
where $\bar x^{\beta}\in \{ \bar x^0,x^1,x^2,x^3 \}$ in a
dielectric.
Like the relativity theory, there is a different conservation law for 
each linear dielectric.
\vskip 0.125in
\par
3) and 4). There is a different set of field equations,
Eqs.~(\ref{EQf4.12})--(\ref{EQf4.15}), with a different timelike coordinate,
$\bar x^0(n)$ for each material.
Consequently, the boundary conditions are indeed non-holonomic.
As we saw above, there is a different principle of dielectric special
relativity and a different conservation law corresponding to, and valid for,
a specific linear isotropic homogeneous material with $\bar x^0(n)$.
Then, without loss of generality, we can drop the right-hand-sides of
Eqs.~(\ref{EQf4.14}) and (\ref{EQf4.14}), leaving us with homogeneous
equations of motion
\begin{subequations}
\begin{equation}
\nabla\times{\bf B} +\frac{\partial {\bf \Pi}}{\partial \bar x^0} =0 
\label{EQg9.18a}
\end{equation}
\begin{equation}
\nabla\cdot{\bf B}= 0
\label{EQg9.18b}
\end{equation}
\begin{equation}
\nabla\times{\bf \Pi}
-\frac{\partial{\bf B}}{\partial \bar x^0}
= 0
\label{EQg9.18c}
\end{equation}
\begin{equation}
\nabla\cdot{\bf \Pi}=0
\label{EQg9.18d}
\end{equation}
\label{EQg9.18}
\end{subequations}
with fields connected by boundary conditions.
The boundary conditions are equivalent to the Fresnel
relations \cite{BIUnpub}
\vskip 0.125in
\par
Finally, there is no evidence for 5) and 6) because the equations are 
carefully derived using small, carefully documented steps from explicit
axioms. 
\vskip 0.125in
\par
We note that Eqs.~(\ref{EQf4.12})--(\ref{EQf4.15}) are not identities 
of the Maxwell--Minkowski equations, Eqs.~(\ref{EQf3.01}), under a
re-parameterization by the direct substitution of
${\bf \Pi}= - n{\bf E}$ and $\bar x^0=ct/n$.
Although that superficially appears to be the case,
the refractive index forms a part of the timelike
coordinate, $\bar x^0$, and defines the flat non-Minkowski material
spacetime.
Specifically, $n$ is not a free parameter as it is in Maxwellian
continuum electrodynamics.
\par
\section{Conclusions}
\par
The macroscopic Maxwell equations lead to a contradiction with
energy--momentum conservation laws.
In this report, we resolved these contradictions by deriving new
equations of motion  for macroscopic fields in a simple linear
dielectric from Lagrangian field theory and by deriving new conservation
laws based on the properties of light in the dielectric.

\end{document}